\documentclass[structabstract]{aa}

\usepackage{graphicx}
\usepackage{txfonts}
\usepackage[cp1250]{inputenc}
\usepackage{natbib}
\newcommand{\tw}{\hbox{TW Dra }}

\newcommand{\ks}{km~s$^{-1}$}

\begin{document}
   \title{Photometric and spectroscopic investigation of TW Draconis}


   \author{M. Zejda \inst{1}, M. Wolf\inst{2}, M. \v{S}lechta \inst{3},
   Z. Mikul\'a\v{s}ek \inst{1,4}, J. Zverko\inst{5}, P. Svoboda\inst{6},
   J. Krti\v{c}ka\inst{1}, J. Jan\'ik\inst{1} and H. Bo\v{z}i\'c\inst{7} }

   \institute{Institute of Theoretical Physics and Astrophysics, Masaryk University,
              Kotl\'a\v{r}sk\'a 2, 61137~Brno, Czech Republic\\
              \email{zejda@physics.muni.cz}
         \and  Astronomical Institute, Faculty of Mathematics and Physics,
               Charles University Prague, V Hole\v{s}ovi\v{c}k\'ach 2, \\
               CZ-180 00 Praha 8, Czech Republic
         \and
            Astronomical Institute, Academy of Science of CR, Ond\v rejov, Czech Republic
        \and
            Observatory and Planetarium of J. Palisa, V\v SB -- Technical University, Ostrava, Czech Republic
        \and
             Astronomical Institute of Slovak Academy of Science, Tatransk\'a Lomnica, Slovakia
        \and
             private observatory, Brno, Czech Republic
        \and
             Hvar Observatory, Faculty of Geodesy, Zagreb University, Zagreb, Croatia}

   \date{Received March, 2011}

  \abstract
{TW~Draconis is one of the best studied Algol-type eclipsing binaries. There is
significant evidence for miscellaneous physical processes between interacting binary
components manifesting themselves by period and light curve changes.   }
   {Obtaining new set of photometric and spectroscopic observations, we analysed them together with the older
   spectroscopic and photometric data to build model of this eclipsing system with respect to observed changes
   of O-C diagram and light curve.}
   {Reduction of new spectra was carried out in the IRAF and SPEFO programs. Radial velocities
   were determined manually using SPEFO, by CCF and from the program KOREL. Orbital
   elements were derived with the FOTEL program and via disentangling with KOREL.  The final combined
   solution was obtained with the programs PHOEBE and FOTEL. }
   {Photometry shows small irregularities in light curves as a results of pulsating of one component
   and spot activity. Using net of KOREL outputs we found the mass ratio q=0.405(3). We confirm the presence
   of stellar matter around the primary. Even after subtraction of ADS 9706B influence, light curve solutions
   show third light in the system. Models in FOTEL and Phoebe are presented.  }
   {\tw is an astrophysically very interesting eclipsing binary. Future combination of interferometry, spectroscopy,
   and photometry is promising. It could definitely confirm the hypothesis of quadruple system for
   \tw\ and explain behaviour of this system in complex.  }


   \titlerunning{Investigation of TW Dra}
   \authorrunning{Zejda et al.}

   \maketitle
%

\section{Introduction}

TW~Draconis (${\alpha}$=15$^{\rm{h}}33^{\rm{m}}51\fs1$, ${\delta}$=63\degr 54\arcmin
26\arcsec (2000.0)) is a well-known and often observed Algol-type eclipsing binary. The
variability of the star was discovered by Cannon (see \citealp{pick10}). The variable
star is an A-component of visual binary ADS~9706. The main light changes with amplitude
in $B$ $\sim$~2.3~mag in primary minimum are caused by 11.5 hours eclipses of the hot
main sequence A8V star by the cooler and fainter K0III giant component.

First light curve solution based on photometric observations was made by \citet{bagl52},
who also mentioned large distortion of the secondary component. However, to obtained good
solutions with low residuals he assumed very low values of limb darkening coefficients
and he obtained also surprisingly low reflection parameters. During further and longer
monitoring of TW~Dra behaviour \citet{walt78} found large changes of orbital period and
changes of the light curve outside the minima, especially around secondary minimum caused
by gas stream and a large hot spot in the target area of it. Observations made by
\citet{bagl52,walt78} were reprocessed by e.g. \citet{giur80} or \citet{alna84}.
\citet{papo84} made several years to obtain light curve solution of the system. They
mentioned larger scatter of the phased light curved but without any discussion in detail.

Although TW~Dra was studied spectroscopically several times since 1919, secondary
component was firstly detected in spectrum by \citet{smit49} and the radial velocity of
both components were measured only by \citet{popp89}. The published values of mass ratio
of the components differs from 0.28 \citep{pear37} to 0.47 \citet{rich94}.

TW~Dra~was also included in several surveys. \citet{sing95} studied X-ray radiation in
five Algol-like stars using ASCA and ROSAT satellites. He obtained first X-ray spektrum
of TW Dra and confirmed previous paper revealing TW Draconis as a X-ray source
\citep{whit83}. \citet{uman91} studied microwave radiation among Algols. They found
radio-fluxes of Algols are generally comparable to them emitting from the RS CVn stars.
During this study in the interval 1984--1989 went the radio-flux of TW Draconis down by
factor 10. In summary although TW Draconis is known and studied the satisfactory complex
solution of this unique system has not been published up to now. \citet{zejd08} published
detailed study of period changes in the system.

The main goal of this study is to find the correct spectroscopic mass ratio of the main
components and to calculate new light curve and radial velocity curve solutions based on
our new measurements.

\section{Observations and reduction}

\subsection{Photometry}

TW~Dra was monitored during photometric campaign in 2001--2007. Almost 50\,000 usable
measurements in \textit{UBVRI} bands were collected (see journal of observations in Tab.
\ref{pozorovatele}). However, only Hvar measurements were transformed into the standard
magnitudes. Suhora observations are only in differential magnitudes as well as all CCD
measurements.

\begin{table}\centering
\caption[]{Journal of photometric observations: Columns: 1. Observer(s); observatory. 2.
Epoch. 3. Diameter/focal length of used telescopes in mm. 4--8. Numbers of measurements
in photometric bands.}\label{pozorovatele}
\renewcommand{\tabcolsep}{2pt}\scriptsize
\vspace{3mm}
\begin{tabular}{llclllll}
\hline Observer(s); place                & Year(s)  & Equipment(s)     & $U$& $B$ & $V$  & $R$ &  $I$  \\
\hline \hline CCD &    &    &    &    &    &    & \\
Br\'{a}t; Pec pod Sn\v{e}\v{z}kou & 2005   & 200/1850   & -  & -   & -    & 154 & - \\
        privat observatory, CZ    &        &            &    &     &      &     &    \\
Hroch; Brno, Masaryk              & 2006   & 620/2780   & -  &  57 &  57  & -   & - \\
university Observ.,CZ             &        &            &    &     &      &     &    \\
Chrastina, Sz\'{a}sz; obs.        & 2005   & 179/1000   & -  & 300 & 300  & 299 & 301 \\
    Hlohovec, Slovakia            &        &            &    &     &      &     &    \\
Svoboda; Brno,                    & 2006-7 & 15/116     &  - & 147 & -    & -   & - \\
      privat observatory,CZ       &        & 35/135     &  - & 688 & -    & -    & 513 \\
Zejda; N. Copernicus              & 2001-7 & 400/1750   &  - & 225 & 6213 & 6089 & 6073 \\
Obs. and planet. Brno, CZ         &        & 40/360     &  - & -   & -    &  900 & 3301  \\
                                  &        & 40/360     &  - & 1662& 639  &  715 & - \\
                                  &        & 200/1000   &  - & 7452& 5017 & 5028 & 5021 \\
\hline
PEP &    &    &    &    &    &    & \\
Zejda, Jan\'{i}k, Bo\v{z}i\'c;    & 2005 & 650/7280   & 136& 136 & 136  & -    & - \\
  Hvar observatory, Croatia       &      &            &    &     &      &      &  \\
Zejda, Jan\'{i}k, Ogloza;         & 2006 & 600/7500   &  - &2547 & 2552 & 2547 & 406 \\
Mt. Suhora obs., Poland           &      &            &    &     &      &      &    \\
\hline
\end{tabular}
\end{table}

The period of the system is changing in at least two modes \citep{zejd08}. The present
photometric data were all obtained in the epoch after the last great period change. Thus
the period in the epoch of our observations is almost constant and light ephemeris in a
linear approximation for this epoch could be used. Mikul\'{a}\v{s}ek's method based on
Principal Component Analysis (hereafter PCA) \citep{miku07} was used to combine all our
photometric measurements on different magnitude scale made in different time intervals to
determine following light ephemeris:
\begin{equation}\label{ephem}
$Pri.Min. = HJD 24\,53558.91888(32) + 2\fd80683211(28)$\cdot E. \label{efe1}
\end{equation}

Despite of our CCD observations lower accuracy, small oscillations were detected on light
curve, in $B$ and $V$ filters, respectively. Pulsation period 0.0519(3) d and
semi-amplitude $\Delta m=10(3)$ mmag \citep{zejd06} found in our \emph{B} measurements
and 0.0501(13)d with $\Delta m=9(2)$ mmag in \emph{B} measurements of \citet{papo84} are
in agreement with \citet{kusa01} (0.0556 d, $\Delta m=2.1$ mmag in $B$) and \citet{kim03}
(0.053 d, $\Delta m=5$ mmag). Change in pulsations amplitude is probably due to
superposition of different pulsation modes as described in RZ~Cas \citep{mkrt08}. Large
scatter in maximum of brightness in eclipsing phased light curve is also caused by
oscillations and their superpositions from different orbital cycles.

\begin{figure}[h]
\centering
\includegraphics[width=0.45\textwidth]{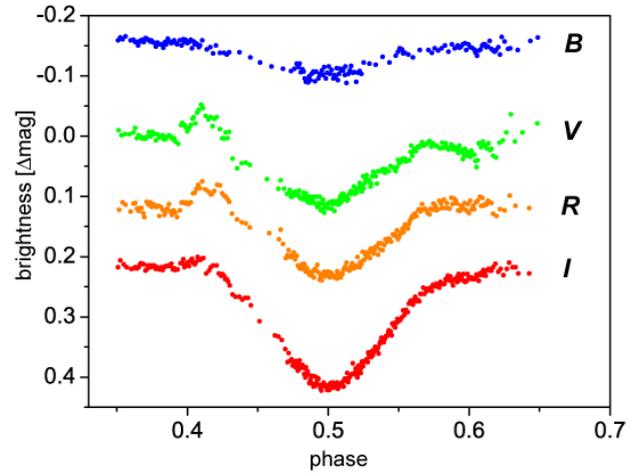}
\caption{Normal light curves around secondary minimum.} \label{seknorm}
\end{figure}

Regardless of confirmed oscillations, deformations of light curve, which persist at least
during several orbital cycles, can be easily seen as well. Our measurements show a small
decrease (0.04~mag in $B$) after the egress from primary eclipse (phase 0.10--0.19)
previously mentioned by \citet{bagl52} and \citet{walt78} and a bump of 0.055~mag ($V$),
0.030 mag ($R$), 0.02 mag ($I$) before the secondary eclipse (phase 0.40--0.43) (see Fig.
\ref{seknorm}).

The most obvious light curve phenomenon is a dependence of the secondary minimum depth on
wavelength. While in $B$ band the secondary minimum is scarcely visible ($\Delta m$=0.063
mag), in $I$ filter is easily detectable ($\Delta m$=0.196 mag) (see Fig. \ref{seknorm}).

\subsection{Spectroscopy}

The spectroscopic observations (46 spectra) were made in JD interval 2453111--2454218
using Coud\'e spectrograph of the 2.0-m Ond\v{r}ejov reflector, equipped with a SITe-005
800x2000 CCD; the spectra cover the H$\alpha$ region from 628.0 to 672.0 nm, with a
linear dispersion of 17 nm\,mm$^{-1}$, two-pixel resolution of 12700, and $S/N$ ranging
up to 478. Spectra with an average signal-to-noise (S/N) ratio lower than 95 in continuum
near H$\alpha$ line were excluded. Finally 37 spectra were analysed (see journal of our
spectroscopic observations in Tab. \ref{rvstab}).

In all cases, the wavelength calibration was based on ThAr comparison spectra taken
before and after the stellar exposure. Mean flatfield images from the same night as the
stellar spectrum were always applied.

The initial reductions to 1-D frames were carried out in IRAF by M\v{S}. The final
reduction, rectification were carried out in SPEFO \citep{horn96,skod96}.

\section{Radial velocities}

We used three methods to obtain RV curves of components of this binary system.
Unfortunately the lines of the secondary component being poorly resolved. The secondary
component was detectable only on some well exposed spectrograms. In SPEFO the RV's were
measured interactively, comparing the direct and flipped images of the line profiles. The
zero point of RV scale was corrected through the use of reliable telluric lines -- see
\citet{horn96} for details.

To obtain RVs we also used the code based on cross-corelation method (see
\citealp{zver07}). The needed synthetic spectra were calculated according to the
parameters from \citet{alna84} by the code SYNTHE (see \citealp{sbor04})\footnote{For the
most recent update of the program and its manual see Sbordone \& Bonifacio 2005,
$http://wwwuser.oat.ts.astro.it/atmos/$ $atlas\_cookbook/Atlas\_Cookbook.html$} and the
code SYNSPEC, which is able to calculate spectra from TLUSTY \citep{lanz07} and ATLAS
model atmospheres.

\begin{figure}
\begin{center}
\includegraphics[width=0.47\textwidth]{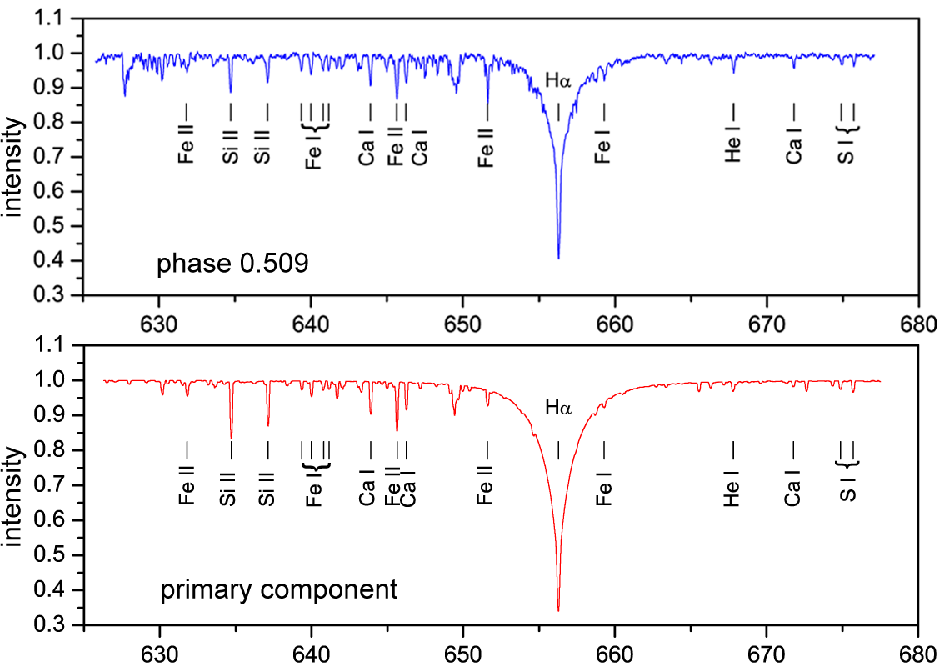}
\includegraphics[width=0.47\textwidth]{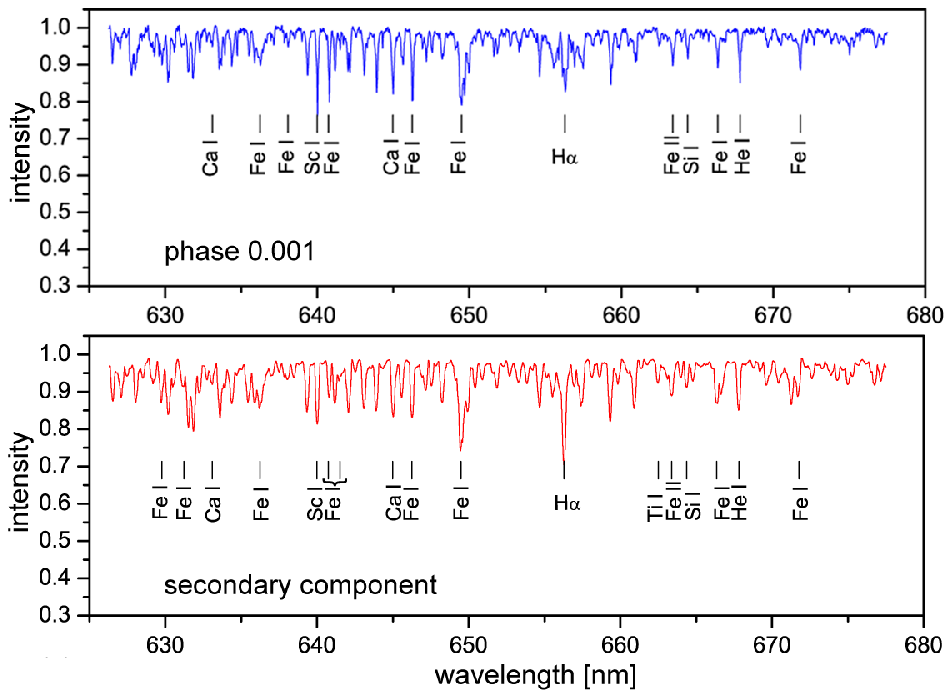}
\caption{Spectrum of \tw~system in the phase 0.509 in comparison with synthetic spectra
of primary component (upper panels). Observed \tw~spectrum in phase 0.001 (containing
secondary component only) as compared with synthetic spectrum of secondary component.
Synthetic spectra are based on Kurucz's model of atmospheres with following parameters
taken from \citet{alna84}: $T_{\rm eff,1}=8300$ K, $\log g_1=3.92$, $v_1\sin i=50$
km~s$^{-1}$, $T_{\rm eff,2}=4800$ K, $\log g_2=3.24$, $v_2\sin i=72$ km~s$^{-1}$, solar
abundance both. It is apparent at observed secondary component spectrum that absorption
line H$\alpha$ is here dimmer than in synthetic spectrum, probably due to an emission in
the system.} \label{spektw}
\end{center}
\end{figure}

Results of the two first methods are shown in Table~ \ref{rvstab}. Finally we used also
the spectral disentangling technique (see relevant section) to obtain radial velocities.

\begin{table*}\begin{center}
\caption[]{Radial velocities in km/s measured in SPEFO (column 6--7), obtained from CCF
(column 8--9)} \vspace{3mm}\label{rvstab} \scriptsize
\begin{tabular}{rccccrrrrc}
No. & $JD_{hel}$ &  Phase  &  S/N  & Weight & $RV_1$ &  $RV_2$ & $RV_1$ &  $RV_2$  & phase\\
\hline
1  & 53111.5121 & 0.6008 &   162.4  &   0.365  &  35.4(1.0) &  -90(4)    &  34.13(3) &              & 0.6008  \\
2  & 53145.3644 & 0.6615 &   227.1  &   0.713  &  53.8(1.8) & -106(3)   &  56.80(3) &  -129.33(49) & 0.6615  \\
3  & 53153.3578 & 0.5093 &   489.3  &   3.311  &   4.8(1.2) &    2(3)      &   5.67(3) &     3.06(30) & 0.5093  \\
4  & 53254.3689 & 0.4969 &   311.6  &   1.343  &   1.1(0.9) &   -4(4)     &   3.10(3) &     7.73(32) & 0.4969  \\
5  & 53454.4151 & 0.7681 &   389.0  &   2.093  &  66.0(1.3) & -103(5)   &  63.05(3) &  -134.91(49) & 0.7681  \\
6  & 53459.4299 & 0.5547 &    98.8  &   0.135  &  16.9(1.3) &  -61(10)   &  19.65(4) &              & 0.5547  \\
7  & 53460.3532 & 0.8837 &   195.6  &   0.529  &  42.3(1.7) &  -98(4)    &  41.46(3) &              & 0.8837  \\
8  & 53461.5080 & 0.2951 &   219.9  &   0.669  & -58.7(1.5) &  131(8)   & -60.71(3) &   133.83(51) & 0.2951  \\
9  & 53463.4444 & 0.9850 &   106.8  &   0.158  &  -1.5(2.7) &  -14(4)    &  9.94(19) &    -7.24(07) & 0.9850  \\
10 & 53463.4912 & 0.0016 &   359.4  &   1.786  &   2.2(1.1) &    2(2)       &  4.50(29) &     2.29(07) & 0.0016  \\
11 & 53464.4606 & 0.3470 &   304.5  &   1.282  & -50.7(1.4) &  108(13)  & -52.39(3) &    99.43(50) & 0.3470  \\
12 & 53465.5142 & 0.7224 &   224.6  &   0.698  &  61.8(1.5) & -110(6)   &  65.20(3) &  -134.91(47) & 0.7224  \\
13 & 53504.4166 & 0.5823 &   262.5  &   0.953  &  28.1(1.0) & -101(5)   &  27.75(3) &              & 0.5823  \\
14 & 53510.4323 & 0.7255 &   305.8  &   1.293  &  59.3(1.5) & -108(5)   &  62.24(3) &  -134.91(53) & 0.7255  \\
15 & 53511.4030 & 0.0713 &   274.5  &   1.042  & -23.1(2.6) &   38(7)    & -23.02(4) &    -4.41(28) & 0.0713  \\
16 & 53516.5391 & 0.9012 &   183.3  &   0.465  &  37.9(1.8) &  -97(4)    &  43.22(4) &              & 0.9012  \\
17 & 53538.3802 & 0.6826 &   243.9  &   0.823  &  57.2(1.8) & -107(3)   &  59.18(4) &  -129.42(48) & 0.6826  \\
18 & 53542.4899 & 0.1468 &   270.9  &   1.015  & -44.7(1.8) &  102(7)   & -49.02(4) &   120        & 0.1468  \\
19 & 53550.4130 & 0.9696 &   137.5  &   0.261  &  11.8(3.7) &   -9(4)     &  30.08(6) &     2.56(19) & 0.9696  \\
20 & 53550.4357 & 0.9777 &   114.9  &   0.183  &   4.1(2.7) &  -13(4)     &  17.54(9) &    -4.63(13) & 0.9777  \\
21 & 53619.3175 & 0.5184 &   386.4  &   2.065  &   7.7(1.3) &  -10(4)     &   9.08(3) &              & 0.5184  \\
22 & 53764.6405 & 0.2932 &   189.4  &   0.496  & -61.3(2.0) &  129(5)   & -55.49(3) &   146.73(52) & 0.2932  \\
23 & 53764.6771 & 0.3062 &   223.0  &   0.688  & -54.2(1.4) &  130(8)   & -54.14(3) &   132.75(55) & 0.3062  \\
24 & 53765.6086 & 0.6381 &   372.3  &   1.917  &  45.7(1.2) & -102(4)   &  49.51(3) &   -93.09(51) & 0.6381  \\
25 & 53846.3238 & 0.3948 &   351.6  &   1.710  & -35.8(0.9) &  109(4)   & -36.33(3) &              & 0.3948  \\
26 & 53860.3570 & 0.3944 &   364.9  &   1.841  & -34.6(1.0) &  103(7)   & -33.91(3) &              & 0.3944  \\
27 & 53991.3876 & 0.0772 &   196.2  &   0.532  & -27.3(1.3) &   61(3)    & -24.87(4) &    -7.60(29) & 0.0772  \\
28 & 54000.4012 & 0.2885 &   330.3  &   1.509  & -62.0(1.4) &  144(5)   & -64.17(4) &   140.44(51) & 0.2885  \\
29 & 54017.3270 & 0.3187 &   175.8  &   0.427  & -56.3(1.1) &  118(7)   & -59.18(4) &   125.60(52) & 0.3187  \\
30 & 54026.3265 & 0.5250 &   211.0  &   0.616  &   8.5(1.2) &   -8(4)   &   9.94(3) &     3.10(28) & 0.5250  \\
31 & 54115.6199 & 0.3378 &   187.9  &   0.488  & -45.4(1.7) &   94(11)   & -45.37(3) &    97.30(49) & 0.3378  \\
32 & 54191.4998 & 0.3718 &   149.7  &   0.310  & -37.8(2.7) &   78(11)   & -40.29(3) &              & 0.3718  \\
33 & 54192.4449 & 0.7086 &   319.6  &   1.413  &  61.7(2.6) &  -94(5)    &  64.13(3) &  -142.46(48) & 0.7086  \\
34 & 54192.4913 & 0.7251 &   191.9  &   0.509  &  58.7(2.9) & -100(8)   &  64.76(3) &  -144.31(49) & 0.7251  \\
35 & 54203.3567 & 0.5961 &   364.4  &   1.836  &  39.8(1.3) & -101(3)   &  38.76(3) &              & 0.5961  \\
36 & 54217.3376 & 0.5772 &   214.2  &   0.634  &  31.6(1.1) &  -92(4)    &  30.62(3) &              & 0.5772  \\
37 & 54218.4308 & 0.9666 &   254.2  &   0.894  &  29.0(2.1) &  -36(6)    &  27.34(5) &    -0.49(21) & 0.9666  \\
\hline
\end{tabular}\\
\vspace{1mm} \scriptsize {Notes: Phases were calculated using ephemeris
\ref{ephem}.}\end{center}
\end{table*}

\subsection{Spectral disentangling}

As a third approach for determination of radial velocity curves the code KOREL was used
\citep{hadr97,hadr04b}. This code enables decomposition of light joint from several
sources using Fourier disentangling with intrinsic line-profile variations. Twelve short
regions in wavelength interval \linebreak 626--677~nm and later on 5 regions from them
joint in one multiple solution were used (see Tab. \ref{regiony}).

\begin{figure}
\centering
\includegraphics[width=0.45\textwidth]{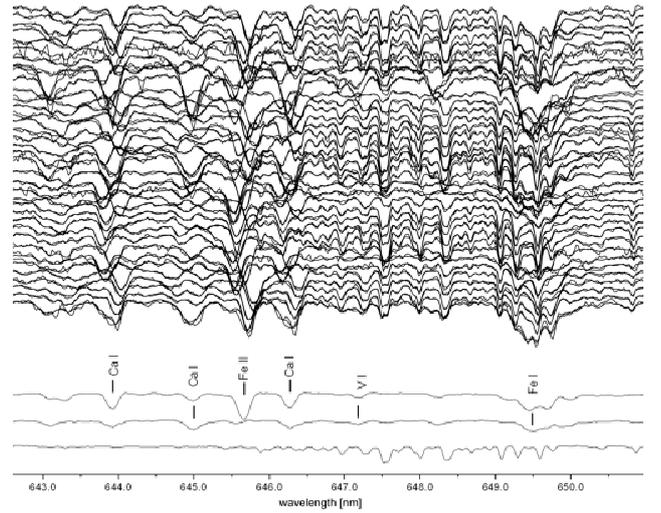}
\caption{KOREL disentangling -- region 642.6--651.0 nm.} \label{Korel1}
\end{figure}

We first measured the $S/N$ for all spectra in the region 661.5--662.0~nm and weighted
them with weights proportional to $(S/N)^2$ and normalized to one for the mean value.
These weights were used in the control file of the auxiliary program PREKOR
\citep{hadr04b} that prepares the input data for KOREL. Firstly we derived the line
strengths of the telluric spectra for the wavelength range 646.5 -- 651.0~nm, which
contains a number of strong water vapour lines and then kept these strengths fixed in all
subsequent solutions with KOREL. We derived the KOREL solutions separately for the
regions shown in the Tab. \ref{regiony}. See an example of done disentagled region in
Fig. \ref{Korel1}.

\begin{figure}[h]
\centering
\includegraphics[width=0.45\textwidth]{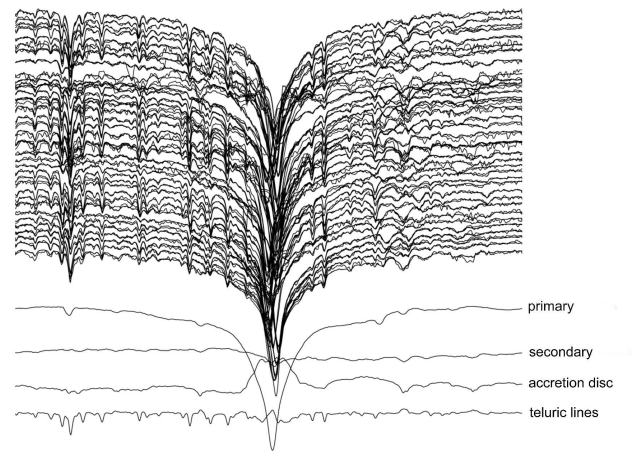}
\caption{KOREL disentangling - H$\alpha$ region. The emission component of spectra was
disentangled as co-rotating object close to primary star.} \label{korelalfa}
\end{figure}

In the process of trial solutions was found that in wavelength interval 650--662~nm the
H$\alpha$ line profile is deformed by presence of emission (see Fig. \ref{korelalfa}),
probably caused by the accretion disc, which is in agreements with preliminary result in
\citet{rich99}. However, this region was not included into following solutions.

\begin{table}
\caption[]{Selected spectral regions used for disentangling.}\vspace{3mm}\label{regiony}
\scriptsize
\begin{center}
\begin{tabular}{cllllll}
No.   & Region     & \multicolumn{2}{l}{Primary comp.} & \multicolumn{2}{l}{Secondary comp.} \\
      &   [nm]     & line &  $\lambda$ & line & $\lambda$    \\
      &            &       &    [nm]   &      &   [nm]      \\
\hline
 1* & 632.50-636.97 &  Si II & 634.7109 & Ca I & 633.0850  \\
   &               &        &          & Fe I & 636.2338    \\
 2 & 633.97-642.69 &  Si II & 634.7109 & Fe I & 636.2338    \\
   &               &  Si II & 637.1371 & Fe I & 638.0743   \\
   &               &  Fe I  & 639.3601 & Fe I & 641.4980   \\
   &               &  Fe I  & 640.0001 &      &            \\
   &               &  Fe I  & 640.8018 &      &            \\
   &               &  Fe I  & 641.1649 &      &            \\
 3 & 634.90-642.76 &  Si II & 637.1371 & Fe I & 636.2338   \\
   &               &  Fe I  & 639.3601 & Fe I & 638.0743   \\
   &               &  Fe I  & 640.0001 & Fe I & 641.4980   \\
   &               &  Fe I  & 640.8018 &      &            \\
   &               &  Fe I  & 641.1649 &      &            \\
 4* & 636.85-641.35 &  Si II & 637.1371 & Fe I & 638.0743   \\
   &               &  Fe I  & 639.3601 &      &    \\
   &               &  Fe I  & 640.0001 &      &    \\
   &               &  Fe I  & 640.8018 &      &    \\
   &               &  Fe I  & 641.1649 &      &    \\
 5 & 641.50-646.03 &  Ca I  & 643.9075 & Ca I & 644.9808 \\
   &               &  Fe II & 645.6383 &      &    \\
 6* & 642.60-651.00 &  Ca I  & 643.9075 & Ca I & 644.9808   \\
   &               &  Fe II & 645.6383 & V  I & 647.1662   \\
   &               &  Ca I  & 646.2567 & Fe I & 649.4980   \\
 7 & 646.50-651.07 &        &          & V I  & 647.1662  \\
   &               &        &          & Fe I & 649.4980   \\
 8 & 650.50-662.96 &  H$\alpha$ & 656.2817 & Ti I & 662.5022  \\
   &               &  Fe I  & 659.2914 &      &  &        \\
 9 & 661.40-666.07 &        &          & Ti I & 662.5022  \\
   &               &        &          & Fe II& 663.3750   \\
   &               &        &          & Si I & 664.3629   \\
10* & 664.00-669.69 &  Fe I  & 667.7987 & Si I & 664.3629  \\
   &               &  He I  & 667.8151 & Fe I & 666.3442   \\
11* & 665.50-670.20 &  Fe I  & 667.7987 & Fe I & 666.3442  \\
   &               &  He I  & 667.8151 &  &    \\
12 & 667.00-671.11 &  Fe I  & 667.7987 &  &    \\
   &               &  He I  & 667.8151 &  &    \\
\hline
\end{tabular}\end{center}
\vspace{1mm} \scriptsize {Notes: The asterisks in the first column shows if the
corresponding spectral region was included in the joint 5-region solution. }
\end{table}

KOREL was able to overcome problems with faint and blended lines of secondary star.
Nevertheless, first solutions (in resolution 2048 bins) gave two mass ratio values with
comparable accuracy; $q_1=0.465$ close to the result of SPEFO, CCF and result of
\citet{popp89} and $q_2=0.408$ close to solution in \citet{lehm09}. This ambiguity can be
found in previously published values of $q$, which differ from 0.28 \citep{pear37} or
0.37 \citep{plas19,holm34} to 0.465 \citep{popp89} or 0.47 \citep{rich94}.

\begin{figure}
\centering
\includegraphics[width=0.48\textwidth]{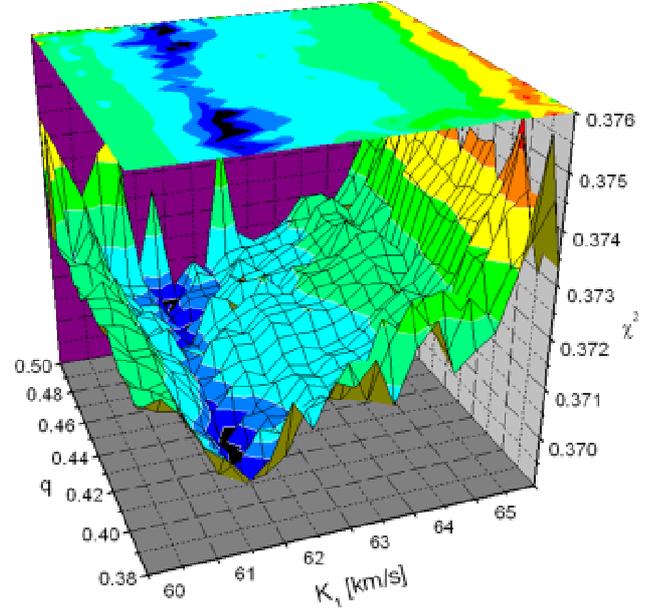}
\caption{Grid of solutions for varying entrance values $q$ and $K_1$ in KOREL, wavelength
interval 642.6--651.0 nm. The global minimum for best solution is the lowest $\chi^2$
value and the darkest place in the diagram.} \label{allnet3d}
\end{figure}

\begin{figure}
\centering
\includegraphics[width=0.48\textwidth]{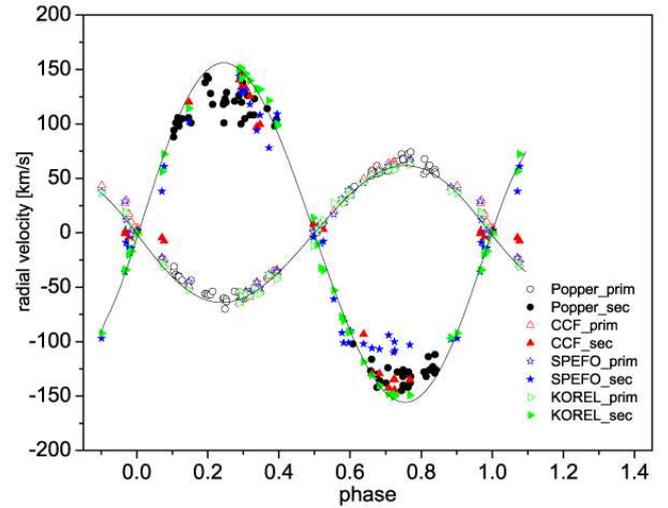}
\caption{Radial velocities determined by SPEFO, CCF and KOREL in comparison with values
taken from
 \citet{popp89}. Phase was calculated according to the found mathematical model of period
 development. For more detail see \citet{zejd08}.} \label{rvall}
\end{figure}

Using region No. 6 (see Tab. \ref{regiony}) we built a grid of solution in resolution 512
bins for input values of $q$ varying from 0.370 to 0.528 with a step 0.002 and
semi-amplitude of primary star radial velocity $K_1$ in interval 61.0 -- 66.8 km/s with a
step 0.2 km/s. During each solution $q$ and $K_1$ parameters were fixed. We obtained 2400
solutions in total. Analogous calculations were repeated for resolutions 256, 512, 1024,
2048 (corresponding to the widths 15.3, 7.6, 3.8 and 1.9 km/s/bin) with smaller steps of
input values: $q$ in 0.380 -- 0.500, step 0.001; $K_1$ in 60.0 -- 66.0, step 0.1. Thus we
obtained 7381 solutions for each resolutions. The global minimum in the space of
solutions was found around $q\approx0.40$ and thus found the final solution for all
chosen spectral intervals (see Fig. \ref{allnet3d}). The resulting mass ratio is
$q=0.405(3)$. The resulting values for the best solutions are summarised in Tab.
\ref{ksol}. Note that disentangling cannot provide information on the systemic velocity.
Its value can only be obtained by a direct measurement of disentangled line profiles.
This way, we found that the systemic velocity of the system is about -2 \ks. The final
radial velocity curves could be compared with other ones in Fig. \ref{rvall}.

\begin{table}
\caption[]{KOREL solutions for the system for selected spectral regions. All epochs are
given in HJD-2400000. Period is fixed $P=2\fd80683211$.}\label{ksol}
\begin{center}
\begin{tabular}{lccccc}
Region & $T_{periastr.}$ & $K_1$ & $K_2$ & $q$ & $\chi^2$ \\
\hline
 1 & 53558.9323 & 64.1 & 160.7 & 0.399 & 0.1996 \\
 2 & 53558.9342 & 62.7 & 155.4 & 0.403 & 0.1865 \\
 3 & 53558.9347 & 62.3 & 158.5 & 0.393 & 0.1771 \\
 4 & 53558.9335 & 63.1 & 160.6 & 0.393 & 0.1987 \\
 5 & 53558.9374 & 63.9 & 151.8 & 0.421 & 0.2212 \\
 6 & 53558.9239 & 62.6 & 154.7 & 0.405 & 0.2191 \\
 7 & 53558.9207 & 62.8 & 158.5 & 0.396 & 0.2009 \\
 9 & 53558.9157 & 63.5 & 156.1 & 0.407 & 0.1663 \\
10 & 53558.9285 & 63.7 & 154.2 & 0.413 & 0.1747 \\
11 & 53558.9212 & 63.3 & 157.1 & 0.403 & 0.1723 \\
12 & 53558.9258 & 66.2 & 157.6 & 0.420 & 0.1555 \\
5 regions & 53558.9219 & 62.7 & 153.7 & 0.408 & 0.2168 \\
\hline
\end{tabular}
\end{center}
\end{table}

\subsection{Initial analysis of RVs}

We also compiled all RVs available in the astronomical literature and used the HEC19
program\footnote{This program which handles data of various forms, together with brief
instructions on how to use it, is freely available on the anonymous ftp server at {\sl
http://ftp.astro.troja.mff.cuni.cz:hec/HEC19\,.}} to derive the heliocentric Julian dates
(HJDs hereafter) for them. In particular, we used the early photographic radial
velocities from the DAO \citep{plas19}, McDonald observatory \citep{smit49} and Lick
observatory obtained by \citet{popp89}. A journal of available spectral observations is
in Table~\ref{jourv}.

\begin{table}
\caption[]{Journal of available spectral observations and RV measurements of TW Dra.}
\label{jourv}
\begin{tabular}{ccrll}
\hline\noalign{\smallskip}
Obs.$^*$&Time interval& No. of &Source\\
       &(HJD$-$2400000)& RVs. \\
\noalign{\smallskip}\hline \hline\noalign{\smallskip}
1&22062.9--22157.7& 14& \citet{plas19}\\
2&32342.6--32377.7& 55& \cite{smit49}\\
3&40290.0--45779.0& 77& \cite{popp89}\\
4&49077.8--49103.8& 104& \cite{rich99}\\
5&49108.7--49139.6& 130& \cite{rich99}\\
6&53335.4--53920.4& 37& this paper   \\
\noalign{\smallskip}\hline
\end{tabular}\vspace{1mm}
{\scriptsize $^*$) In column {\sl ``Obs."}, individual instruments are identified by
numbers: 1... Dominion Astrophysical Observatory 1.83-m reflector, pg plates; 2...2-m
McDonald telescope, McDonald observatory, Cassegrain spectrograph, pg plates; 3...3-m
Shane telescope at the Lick observatory, pg plates; 4... 1.5 m National Solar Observatory
(NSO) McMath-Pierce Main Telescope at Kitt Peak, echelle CCD spg.; 5... 0.9 m Coud\'e
Feed Telescope at Kitt Peak National Observatory (KPNO), CCD spg.; 6... Ond\v{r}ejov 2-m
telescope, coud\'e CCD spg.}
\end{table}

All trial solutions of the RV were derived with the FOTEL program \citep{hadr90,hadr04a};
for the results see Tab. \ref{fotelrv}.

\begin{table}
\caption[]{Journal of RVs curves solutions using FOTEL.}\vspace{3mm}\label{fotelrv}
\begin{center}
\begin{tabular}{lllll}
 Ref. & $P$ [day] & $T_0$ (JD-2400000) & $K_1$ [km/s] & ~~~$q$   \\
\hline
 1  & 2.80654 & 22139.605(14) & 65.1(16) &    \\
 2  & 2.806 & 32341.891(17) & 62.9(21) &     \\
 3  & 2.80686656 & 42258.483(9) & 63.0(6) & 0.465(7)   \\
 4  & 2.80686656 & 42258.501(8) & 63.3(6) &     \\
 5  & 2.80683211 & 53558.916(12) & 60.9(10) & 0.474(18)   \\
 6  & 2.80683211 & 53558.938(16) & 62.7(9) & 0.468(23)   \\
 7  & 2.80683211 & 53558.938(8)  & 62.6(12) & 0.441(13)   \\
 8  & 2.80683211 & 53558.9219(15) & 62.7(6) & 0.405(3)   \\
\hline
\end{tabular}
\end{center}
\vspace{1mm} \scriptsize{Notes: In column {\sl ``Ref"}, the RVs sources are identified by
numbers: 1...\citet{plas19}, 2... \citet{smit49}, 3... \citet{popp89}, 4...
\citet{popp89} bv observations, 5... this paper, SPEFO method, 6... this paper, CCF
method, 7... this paper, CCF method, second template, 8... this paper, KOREL.}
\end{table}

\section{B-component of visual binary ADS9706B}

TW Draconis is A-component of visual binary O$\Sigma$ 299 = ADS 9706. The B-component is
the star HD 140512=TYC 4184~61~2 9.987 mag  (VT) only 3.3" far away from TW Dra. ADS9706B
itself has been photometrically observed only rarely. The most accurate measurements were
done by Hipparcos satellite, where A-component (\tw) is $H_{\rm p}=7.517\pm0.006$ mag and
ADS9706B $H_{\rm p}=9.887\pm0.047$ mag. Both components were measured together during our
photometric measurements. To subtract the influence of B-component we obtained also three
spectroscopic observations of this star. Two H$\alpha$ region spectra of B-component from
Ond\v{r}ejov observatory and one echelle spectrum from Tautenburg \citep{mkrt07} were
obtained. Using equivalent width of chosen spectral lines and comparison to the synthetic
spectra, the spectral type of ADS9706B as F5V-G0V was determined. Measured $H_p$
magnitudes were transformed into \textit{V} magnitudes in Johnson system, according to
\citet{harm98}. The result for G0V spectral type was closer to our observational data.
Thus, this spectral type and consequent parameters were used to subtract ADS9706B
influence from our \tw~photometry.

Two-colours diagram was created using decomposed brightnesses of both components and
colour indices of stars in the close vicinity of \tw~(taken from the photometric database
\citep{merm07}). Measured values were compared to theoretical models from \citet{gola74,
cram84}. Considering this diagram, small distance of \tw~$d=122\pm15$ pc \citep{perr97}
and high galactic latitude (45°), interstellar extinction was omitted. According to
\citet{cham92}, visual B-component could be a physical component of the system (along
with \tw). Using spectral type (G0V) of ADS9706B, distance 118 pc was obtained. This
value is in agreement with the distance of \tw~itself and thus supports Chambliss's
conclusion. Though first astrometric measurements of ADS9706 were done in 1843, there is
no significant change of position angle or angular distance up to now.

\section{Combined light-curve and orbital solution}

For full-fledged modelling of all available data we used the code FOTEL \citep{hadr04a}
and the PHOEBE program version 0.29c \citep{prsa05,prsa06} with the WD back-end
\citep{wils71}. To improve the $S/N$ of photometric data, we created 1000 normal points
in \textit{BVRI} colors; in U band we used all 136 original individual measurements. We
excluded all observations from obtained with higher scatter in case of either a technical
problem or bad weather. RVs for the final solutions were taken from the KOREL solution.
We fitted the model to photometric and RV data simultaneously. The limb-darkening
coefficients for the square-root law were interpolated from \citet{vanh93} tables.

\subsection{FOTEL}
Photometric measurements of \tw~from \citet{bagl52, walt78, papo84} were resolved as
well. It is necessary to consider third light in the system to improve to quality and
accuracy of results. Presence of the third light is in agreement with previously
published solutions.

Our own measurements were split into two datasets -- photoelectric photometry from Hvar
observatory transformed into the standard \textit{UBV} magnitude and all (mainly CCD)
observations in differential \textit{UBVRI} magnitudes. Hvar measurements are
approximately the same amount as Baglow's ones, but unfortunately not covering secondary
minimum. Thus, solution with the third light fixed was calculated in this case only.
Initial light ephemeris is: $P$=2.80683211~d, $M_0=53558.91888$. The period $P$, mass
ratio $q=0.405$ and the eccentricity $e=0$ were fixed. Initial values of the other
parameters ($r_1, r_2, i, q, K_1$) were adopted from \citet{bagl52}. Limb darkening
coefficients were interpolated according to tables from \citet{vanh93}.

The second, CCD datasets include normal light curves in \textit{BVRI} bands, each of them
with 1000 points. Hereafter only these normal CCD light curves were utilised, however
initial parameters were the same as described above. Resulting parameters are given in
the first column of Tab. \ref{sumtab}. Solutions based on both datasets are in very good
agreement.

\begin{figure}
\begin{center}
\includegraphics[width=0.24\textwidth]{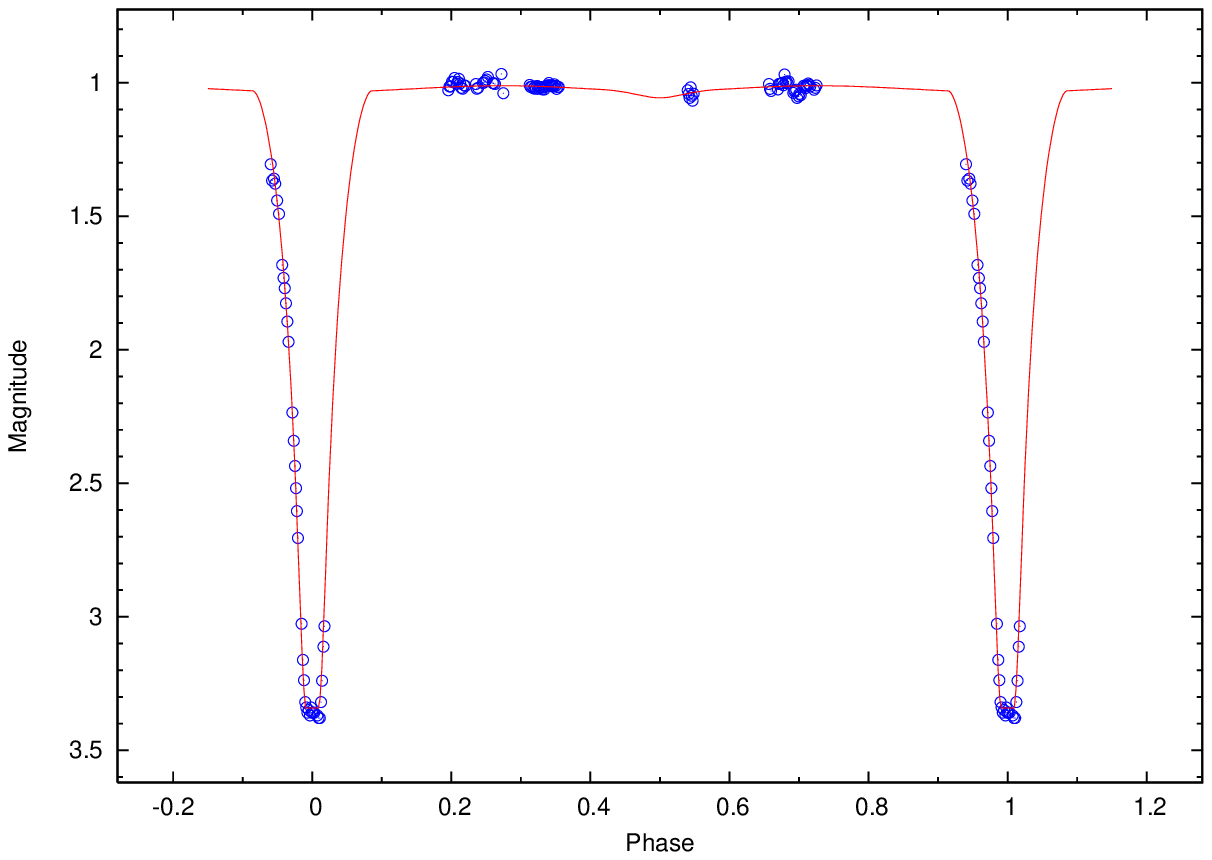}
\includegraphics[width=0.24\textwidth]{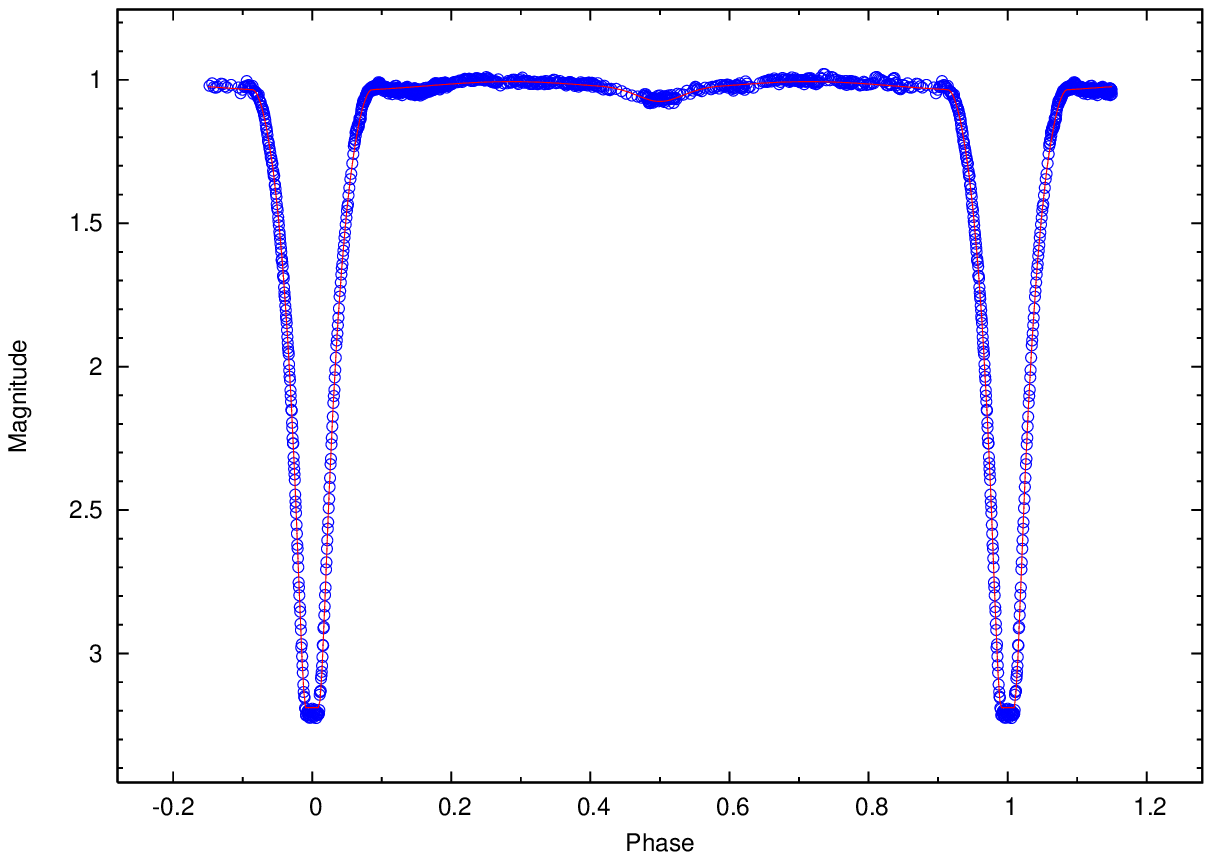}
\includegraphics[width=0.24\textwidth]{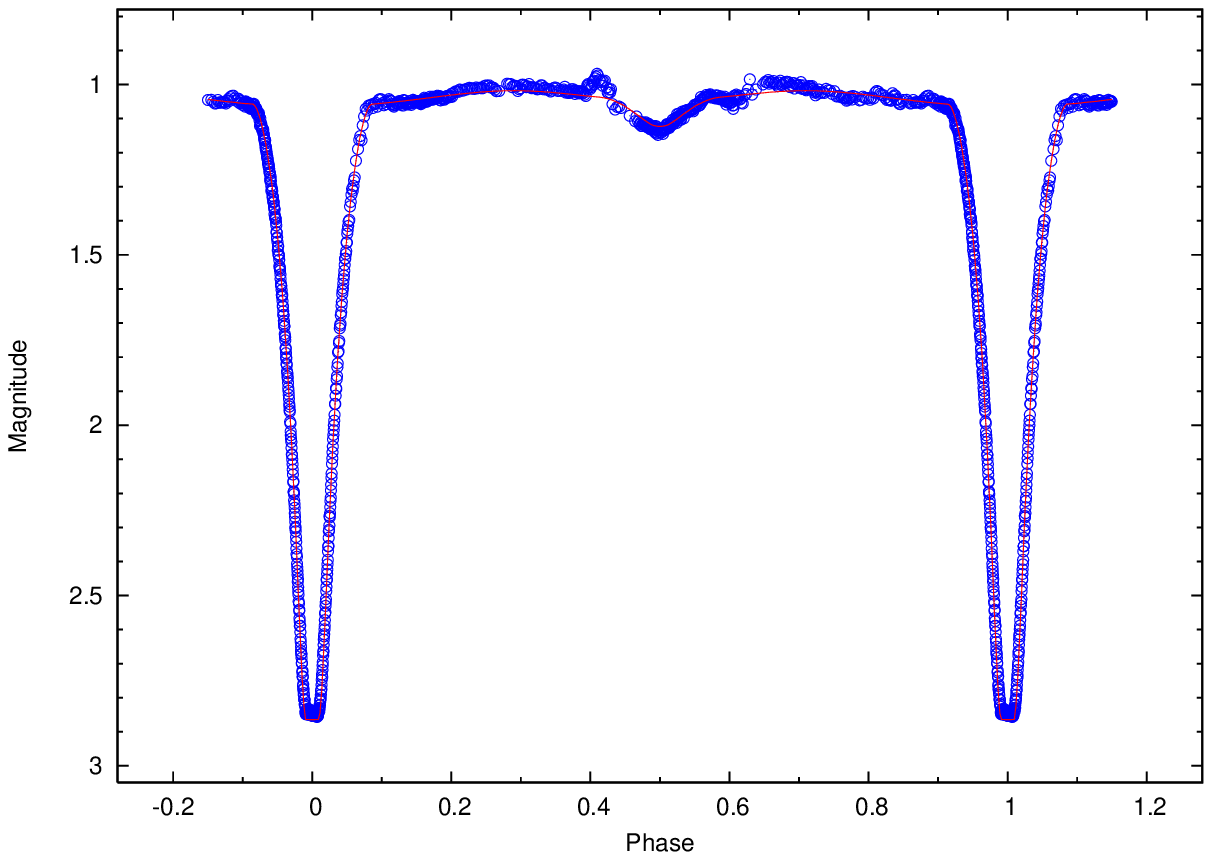}
\includegraphics[width=0.24\textwidth]{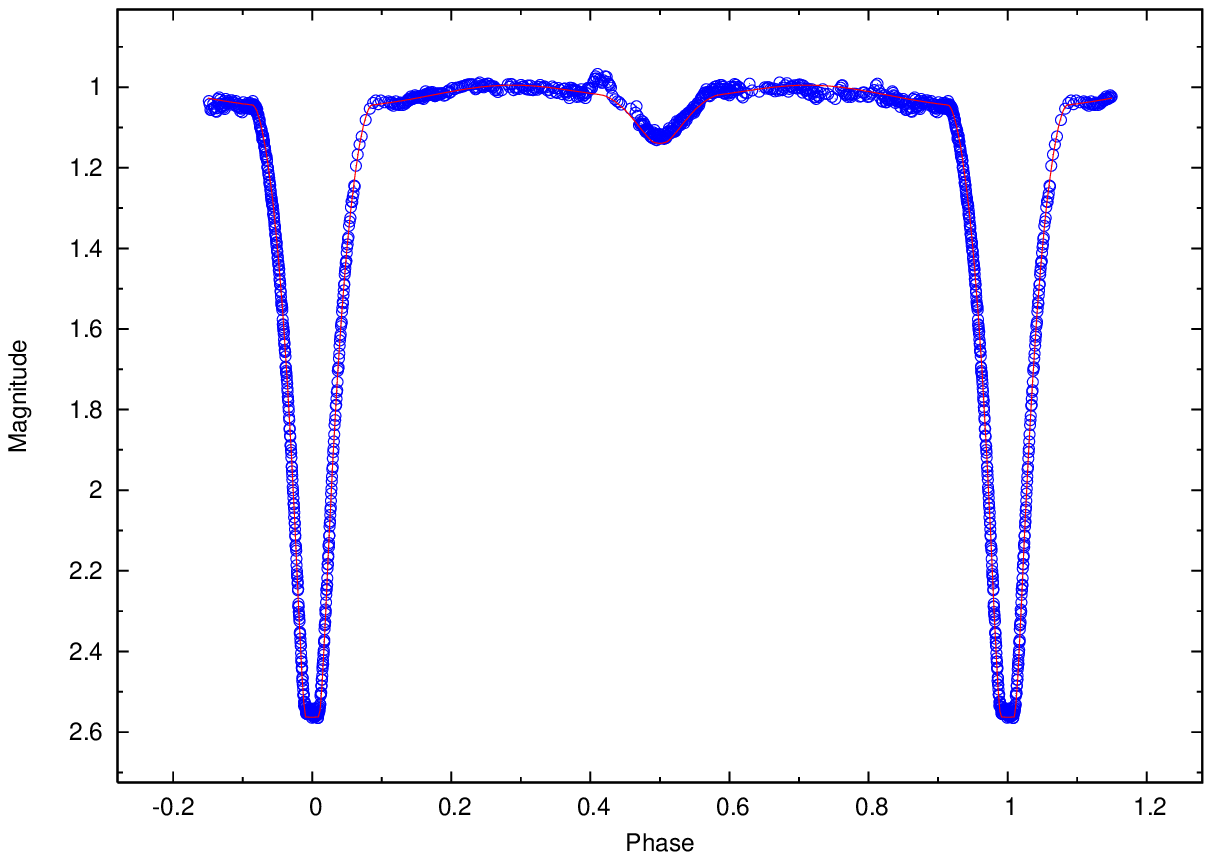}
\includegraphics[width=0.24\textwidth]{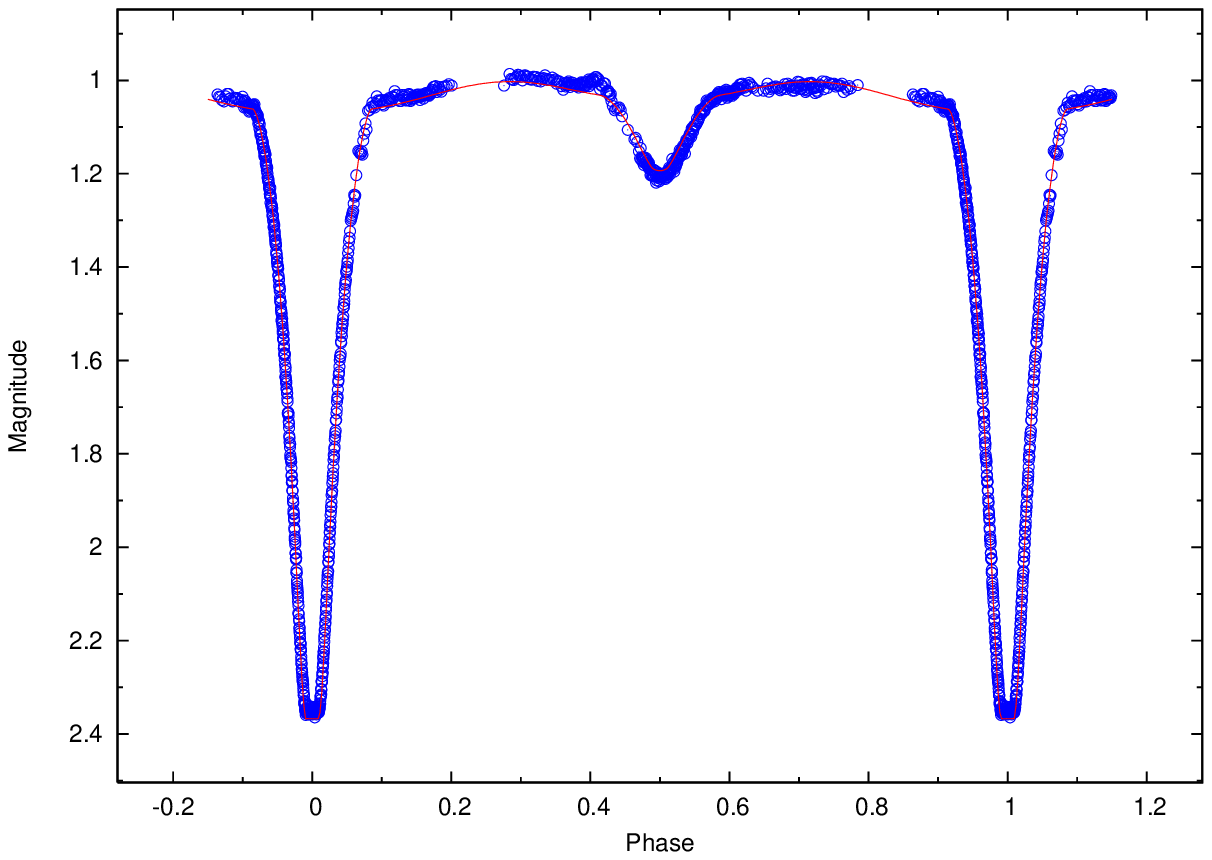}
\includegraphics[width=0.24\textwidth]{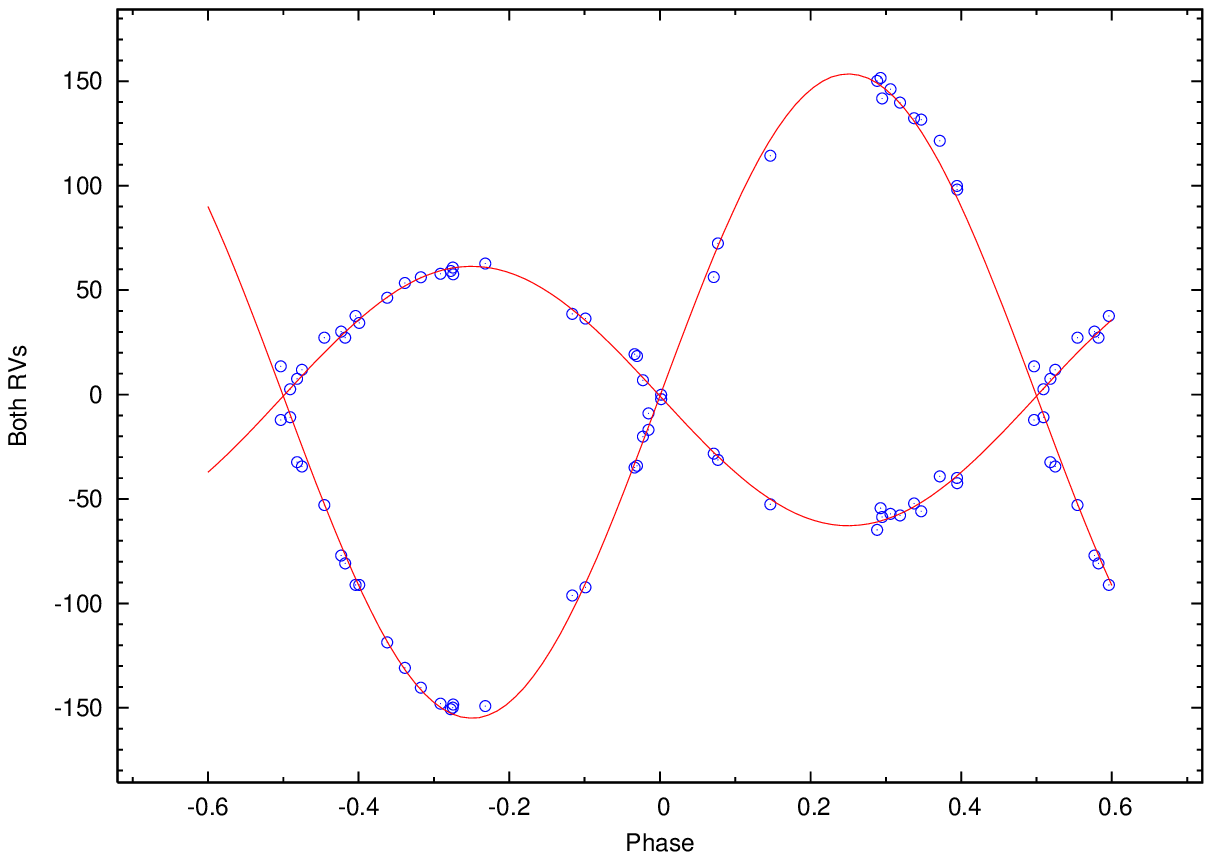}
\caption{Light curves of \tw~together with solution from PHOEBE in \textit{UBVRI} bands
\emph{(from upper left)}. Radial velocity curve with the solution from PHOEBE
\emph{(bottom right)}.} \label{RVphoebe}
\end{center}
\end{figure}

\subsection{PHOEBE}
According to experience with our datasets in FOTEL we have proceeded solutions with non
zero third light in PHOEBE only. First, radial velocities were used to improve initial
values of parameters associated with spectroscopy with following results: semi-major axis
$a=12.0124(3)$ R$_{\odot}$, mass ratio $q=0.4026(11)$, inclination $i=87.13(5) ^\circ$,
system velocity $\gamma$=-0.713(23) km~s$^{-1}$. The $a, q, \gamma$ parameters were then
fixed.

Input parameters were the same for both PHOEBE and FOTEL analysis, except above mentioned
$a, q, i, \gamma$ parameters. In contrast to FOTEL, temperatures of components were
fitted in PHOEBE. Due to presence of third light it was impossible to use reliably either
colour constraining method or the depth of minima for at least temperature ratio
determination. Thus, temperature of primary star was estimated according to spectral type
A5V $T_1=8180$~K \citep{cox99}. Based on the range of possible spectral types of
secondary component K2III, K0III, K2V \citep{popp89,yoon94,samu09}, corresponding
temperature of secondary star is in the interval 4390--4830~K \citep{cox99}. Solutions
for the mode 2 (detached systems) and mode 5 (algols) were proceeded and resulting
parameters are listed in 2nd and 3rd columns of Tab. \ref{sumtab}, showing similar
results with only small differences. Thus the secondary component fills its Roche lobe or
is only slightly below the Roche limit.

\begin{table}
\normalsize \setlength{\tabcolsep}{3pt} \caption{Table of  \tw~light curve solution based
on photometric data obtained during the campaign 2001-2007 (all with non zero third
light). Period was fixed $P=2.80683211$ d. Fixed parameters are marked by an asterisk.
Full version of the table is available in \citet{zejd08b}.}
\label{sumtab}
\renewcommand{\tabcolsep}{2pt}\begin{center}
\begin{tabular}{l|lccc}
             & band& FOTEL &  \multicolumn{2}{c}{PHOEBE} \\
             &     & including &  mode 2  & mode 5  \\
             &     & 3rd light &  (detached) & (algols) \\
\hline
$T_0$        &         &                  &                 &                  \\
$(53558.+)$  &         &   .91894(4)      &    .92013(25)   &    .91879(6) \\
$r_1$        &         &   0.2166(5)      &    0.2139       &    0.2106     \\
$r_{1,pole}$ &         &   0.2151         &    0.2127       &    0.2096(14)    \\
$r_{1,point}$&         &   0.2180         &    0.2157       &    0.2124(17)    \\
$r_{1,side}$ &         &   0.2166         &    0.2142       &    0.2110(15)    \\
$r_{1,back}$ &         &                  &    0.2153       &    0.2120(16)    \\
$r_2$        &         &   0.3144(3)      &    0.3047       &    0.3030        \\
$r_{2,pole}$ &         &   0.2918         &    0.2839       &    0.2830        \\
$r_{2,point}$&         &   0.3529         &                 &    0.4077        \\
$r_{2,side}$ &         &   0.3077         &    0.2961       &    0.2950        \\
$r_{2,back}$ &         &                  &    0.3294       &    0.3277        \\
$i$[$^\circ$]&         &   86.57(6)       &    87.30(8)     &    87.13(3)      \\
a [R$_{\odot}$]&       &   12.084         &    12.0124*     &    12.0124*      \\
$M_1$ [M$_{\odot}$]&   &   2.14           &    2.112        &    2.112(53)     \\
$M_2$ [M$_{\odot}$]&   &   0.87           &    0.850        &    0.850(21)     \\
$q$          &         &   0.405          &    0.4026*      &    0.4026*       \\
$\gamma$     &         &   -0.7116        &    -0.713*      &    -0.713*       \\
$\Omega_1$   &         &                  &    5.095(11)    &    5.166(15)     \\
$\Omega_2$   &         &                  &    2.679(1)     &    2.680*        \\
log $g_1$    &         &                  &    3.940        &    3.960         \\
log $g_2$    &         &                  &    3.240        &    3.240         \\
$T_1 [K] $   &         &                  &    8180*        &    8180*         \\
$T_2 [K] $   &         &                  &    4437(18)      &    4407(23)     \\
$M_{bol,1} [mag]$ &    &                  &    1.260        &    1.260         \\
$M_{bol,2} [mag]$ &    &                  &    3.120        &    3.160         \\
             &         &                  &                 &                  \\
$\Sigma$(O-C)$^2$&      U  &   0.0249         &    0.0185       &    0.0213        \\
             &      B  &   0.0140         &    0.0102       &    0.0099        \\
             &      V  &   0.0149         &    0.0125       &    0.0124        \\
             &      R  &   0.0138         &    0.0122       &    0.0109        \\
             &      I  &   0.0117         &    0.0102       &    0.0107        \\
$\Sigma$(O-C)$^2$ RV&      &   0.8284         &    1.7189       &    1.7181        \\
\hline
\end{tabular}
\end{center}
\end{table}

\section{Conclusions}

Using rich photometric material we found new linear ephemeris (in agreement with previous
period study \citep{zejd08}). Although this study was not aimed to detailed study of
detailed photometric behaviour, we confirmed pulsation of one component in our data as
well as in previous set od data from \citet{papo84}. There are also small irregularities
in light curves caused by spot activity on surfaces of components or in surrounding
stellar matter.

Spectroscopic part concentrated on several tasks.  Radial velocities were obtained from
new spectra applying three methods (SPEFO, CCF, KOREL); radial velocity curves for both
components were created and mass ratio of component was found with the value $q=0.405(3)$
(KOREL, FOTEL) and $q=0.4026(11)$ (Phoebe solution), respectively. Using a disentangling
technique a presence of interstellar material in the system presumably in a form of
accretion disc was confirmed. Detailed study using set of old yet unprocessed spectra
\citet{rich99} as well as new huge collection of spectra from Tautenburg \citep{mkrt08b}
could help to study the presence of circumstellar matter in the system and its evolution
in time.

The first obtained spectra of visual component ADS9706B of \tw~served for its spectral
type determination (G0V). This result supports previously published hypothesis that
ADS9706B is not only visual but even a physical member of the system \citep{cham92}.
However, \citet{zejd08} found on the base on light time effect appeared in O-C residuals
another component which is too faint to exert influence upon the light of two basic
component and above mentioned third component (ADS9706B). Thus \tw~itself is a basic
binary in a quadruple system ((1+1)+1)+1. The spectral type G0V also allows to subtract
the influence of ADS9706B from total light of visual pair ADS9706. Nevertheless new more
precise spectroscopic and photometric measurements of B-component are highly desirable.

\begin{figure}
\begin{center}
\includegraphics[width=0.45\textwidth]{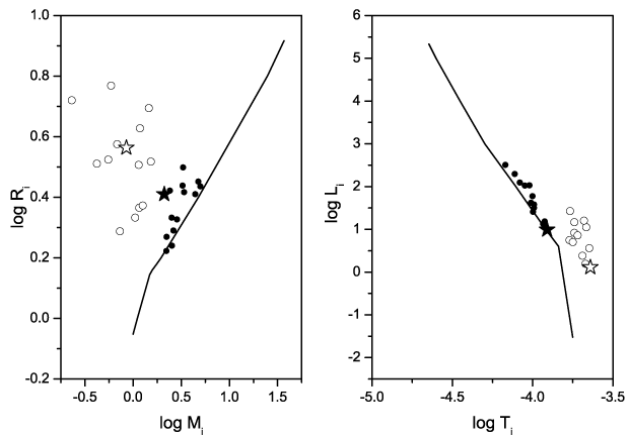}
\caption{The comparison of radii $R_i$, masses $M_i$, luminosities $L_i$ and temperatures
$T_i$ of chosen algols according to \citet{hild01}. Primary components (full circles),
secondary (empty circles) in mass--radius plane (left) and in~HR~diagram (right) together
with ZAMS lines. Asterisks indicate a position of \tw~components.} \label{modelycomp}
\end{center}
\end{figure}

Radial velocity and light curve solutions were done in FOTEL using entire (historical and
current) photometric and spectroscopic observations. New photometric and spectroscopic
observations were processed in PHOEBE as well. The parameters from both FOTEL and PHOEBE
are in very good agreement. \tw belongs to small numbers of semidetached eclipsing
binaries with well known basic parameters. The nowadays primary components of these
systems (e.g. KO Aql, S Cnc, RZ Cas, TV Cas, U CrB, AI Dra, S Equ, TZ Eri, AF Gem, TT
Hya, BP Mus, AT Peg, HU Tau, TX UMa and others) obtaining material from the secondary
present themselves like main sequence stars, while the the same age secondaries are
bigger and more radiative. Obtained results of \tw\ correspond to other algols with known
accurate parameters (see Fig. \ref{modelycomp}).

\begin{acknowledgements}
We acknowledge the use of the programs FOTEL, KOREL, and PREKOR, made available by their
author, Petr~Hadrava, the use of HEC 19 made by Petr Harmanec and also use of PHOEBE made
by Andrej Pr\v{s}a and his group. We appreciate the help and valuable notes of Petr
Hadrava, Petr Harmanec and Volkan Baki\c{s}. This research was supported from the grants
GA~\v{C}R 205/04/2063, 205/06/0217, P209/10/0715 of the Czech Science Foundation,
MEB050819, ME08038, MEB080832 Ministry of education, grant of Academy of Science GAAV
IAA301630901, grant of Slovak VEGA 2/0074/09. We acknowledge the use of the electronic
bibliography maintained by NASA/ADS system and by the CDS in Strasbourg.
\end{acknowledgements}



\begin{thebibliography}{49}
\bibitem[{{Al-Naimiy} \& {Al-Sikab}(1984)}]{alna84}
{Al-Naimiy}, H. M.~K. \& {Al-Sikab}, A.~O. 1984, \apss, 103, 115

\bibitem[{{Baglow}(1952)}]{bagl52}
{Baglow}, R.~L. 1952, Publications of the David Dunlap Observatory, 2

\bibitem[{{Chambliss}(1992)}]{cham92}
{Chambliss}, C.~R. 1992, \pasp, 104, 663

\bibitem[{{Cox}(1999)}]{cox99}
{Cox}, A.~N. 1999, {Allen's astrophysical quantities}, ed. A.~N. {Cox}

\bibitem[{{Cramer}(1984)}]{cram84}
{Cramer}, N. 1984, \aap, 132, 283

\bibitem[{{Giuricin} {et~al.}(1980){Giuricin}, {Mardirossian}, \&
  {Predolin}}]{giur80}
{Giuricin}, G., {Mardirossian}, F., \& {Predolin}, F. 1980, \apss, 73, 389

\bibitem[{{Golay}(1974)}]{gola74}
{Golay}, M., ed. 1974, Astrophysics and Space Science Library, Vol.~41,
  {Introduction to astronomical photometry}, 79--80

\bibitem[{{Hadrava}(1990)}]{hadr90}
{Hadrava}, P. 1990, Contributions of the Astronomical Observatory Skalnat\'e
  Pleso, 20, 23

\bibitem[{{Hadrava}(1997)}]{hadr97}
{Hadrava}, P. 1997, \aaps, 122, 581

\bibitem[{{Hadrava}(2004{\natexlab{a}})}]{hadr04a}
{Hadrava}, P. 2004{\natexlab{a}}, Publ. Astron. Inst. Acad. Sci. Czech Rep.,
  92, 1

\bibitem[{{Hadrava}(2004{\natexlab{b}})}]{hadr04b}
{Hadrava}, P. 2004{\natexlab{b}}, Publ. Astron. Inst. Acad. Sci. Czech Rep.,
  92, 15

\bibitem[{{Harmanec}(1998)}]{harm98}
{Harmanec}, P. 1998, \aap, 335, 173

\bibitem[{{Hilditch}(2001)}]{hild01}
{Hilditch}, R.~W. 2001, {An Introduction to Close Binary Stars}, ed. R.~W.
  {Hilditch}

\bibitem[{{Holmberg}(193)}]{holm34}
{Holmberg}, E. 193, Meddelanden Lunds Astron. Observ., Ser. II, Band VIII, 71,
  1

\bibitem[{{Horn} {et~al.}(1996){Horn}, {Kub\'at}, {Harmanec}, {Koubsk\'y},
  {Hadrava}, {\v{S}imon}, {\v{S}tefl}, \& {\v{S}koda}}]{horn96}
{Horn}, J., {Kub\'at}, J., {Harmanec}, P., {et~al.} 1996, \aap, 309, 521

\bibitem[{{Kim} {et~al.}(2003){Kim}, {Lee}, {Kwon}, {Youn}, {Mkrtichian}, \&
  {Kim}}]{kim03}
{Kim}, S.-L., {Lee}, J.~W., {Kwon}, S.-G., {et~al.} 2003, \aap, 405, 231

\bibitem[{{Kusakin} {et~al.}(2001){Kusakin}, {Mkrtichian}, \&
  {Gamarova}}]{kusa01}
{Kusakin}, A.~V., {Mkrtichian}, D.~E., \& {Gamarova}, A.~Y. 2001, Information
  Bulletin on Variable Stars, 5106, 1

\bibitem[{{Lanz} \& {Hubeny}(2007)}]{lanz07}
{Lanz}, T. \& {Hubeny}, I. 2007, \apjs, 169, 83

\bibitem[{{Lehmann} {et~al.}(2009){Lehmann}, {Tkachenko}, \&
  {Mkrtichian}}]{lehm09}
{Lehmann}, H., {Tkachenko}, A., \& {Mkrtichian}, D.~E. 2009, Communications in
  Asteroseismology, 159, 45

\bibitem[{{Lehmann} {et~al.}(2008){Lehmann}, {Tkachenko}, {Tsymbal}, \&
  {Mkrtichian}}]{mkrt08b}
{Lehmann}, H., {Tkachenko}, A., {Tsymbal}, V., \& {Mkrtichian}, D.~E. 2008,
  Communications in Asteroseismology, 157, 332

\bibitem[{{Mermilliod} {et~al.}(2007){Mermilliod}, {Hauck}, \&
  {Mermilliod}}]{merm07}
{Mermilliod}, J.-C., {Hauck}, B., \& {Mermilliod}, M. 2007,
  http://obswww.unige.ch/gcpd/gcpd.html

\bibitem[{{Mikul\'{a}\v{s}ek}(2007)}]{miku07}
{Mikul\'{a}\v{s}ek}, Z. 2007, Astron and Astroph. Transact., 26, 63

\bibitem[{{Mkrtichian} \& {Lehmann}(2007)}]{mkrt07}
{Mkrtichian}, D.~E. \& {Lehmann}, H. 2007, private communication

\bibitem[{{Mkrtichian} {et~al.}(2008){Mkrtichian}, {Lehmann}, {Lee}, \&
  et~al}]{mkrt08}
{Mkrtichian}, D.~E., {Lehmann}, H., {Lee}, B.~C., \& et~al. 2008, in Proceed.
  of "The 1st Thailand and Korea Joint Workshop on Stellar Astrophysics", Chang
  Mai Univ. Press, Chang Mai, Thailand, 49

\bibitem[{{Papou{\v s}ek} {et~al.}(1984){Papou{\v s}ek}, {Tremko}, \& {Veste{\v
  s}n{\'{\i}}k}}]{papo84}
{Papou{\v s}ek}, J., {Tremko}, J., \& {Veste{\v s}n{\'{\i}}k}, M. 1984, Folia
  Fac.~Sci.~Nat.~Univ.~Purkynianae Brun., Phys., Tomus 25, Opus 4, 64 pp., 25

\bibitem[{{Pearce}(1937)}]{pear37}
{Pearce}, J.~A. 1937, in Publications of the American Astronomical Society,
  Vol.~9, Publications of the American Astronomical Society, 131--+

\bibitem[{{Perryman} \& {ESA}(1997)}]{perr97}
{Perryman}, M.~A.~C. \& {ESA}, eds. 1997, ESA Special Publication, Vol. 1200,
  {The HIPPARCOS and TYCHO catalogues. Astrometric and photometric star
  catalogues derived from the ESA HIPPARCOS Space Astrometry Mission}

\bibitem[{{Pickering}(1910)}]{pick10}
{Pickering}, E.~C. 1910, Harvard College Obs. Circ., 159, 3

\bibitem[{{Plaskett}(1919)}]{plas19}
{Plaskett}, J.~S. 1919, Publications of the Dominion Astrophysical Observatory
  Victoria, 1, 137

\bibitem[{{Popper}(1989)}]{popp89}
{Popper}, D.~M. 1989, \apjs, 71, 595

\bibitem[{{Pr{\v s}a}(2006)}]{prsa06}
{Pr{\v s}a}, A. 2006

\bibitem[{{Pr{\v s}a} \& {Zwitter}(2005)}]{prsa05}
{Pr{\v s}a}, A. \& {Zwitter}, T. 2005, \apj, 628, 426

\bibitem[{{Richards} \& {Albright}(1994)}]{rich94}
{Richards}, M.~T. \& {Albright}, G.~E. 1994, in Interacting binary stars: a
  symposium San Diego State Univ.,13-15 July 1993., ed. A.~{Shafter}, Vol.~56,
  393

\bibitem[{{Richards} \& {Albright}(1999)}]{rich99}
{Richards}, M.~T. \& {Albright}, G.~E. 1999, \apjs, 123, 537

\bibitem[{{Samus} {et~al.}(2009){Samus}, {Durlevich}, \& {et al.}}]{samu09}
{Samus}, N.~N., {Durlevich}, O.~V., \& {et al.} 2009, VizieR Online Data
  Catalog, 1, 2025

\bibitem[{{Sbordone} {et~al.}(2004){Sbordone}, {Bonifacio}, {Castelli}, \&
  {Kurucz}}]{sbor04}
{Sbordone}, L., {Bonifacio}, P., {Castelli}, F., \& {Kurucz}, R.~L. 2004,
  Memorie della Societa Astronomica Italiana Supplement, 5, 93

\bibitem[{{Singh} {et~al.}(1995){Singh}, {Drake}, \& {White}}]{sing95}
{Singh}, K.~P., {Drake}, S.~A., \& {White}, N.~E. 1995, \apj, 445, 840

\bibitem[{{\v{S}koda}(1996)}]{skod96}
{\v{S}koda}, P. 1996, in ASP Conf. Ser. 101: Astronomical Data Analysis
  Software and Systems V, 187--189

\bibitem[{{Smith}(1949)}]{smit49}
{Smith}, B. 1949, \apj, 110, 63

\bibitem[{{Umana} {et~al.}(1991){Umana}, {Catalano}, \& {Rodono}}]{uman91}
{Umana}, G., {Catalano}, S., \& {Rodono}, M. 1991, \aap, 249, 217

\bibitem[{{van Hamme}(1993)}]{vanh93}
{van Hamme}, W. 1993, \aj, 106, 2096

\bibitem[{{Walter}(1978)}]{walt78}
{Walter}, K. 1978, \aaps, 32, 57

\bibitem[{{White} \& {Marshall}(1983)}]{whit83}
{White}, N.~E. \& {Marshall}, F.~E. 1983, \apjl, 268, L117

\bibitem[{{Wilson} \& {Devinney}(1971)}]{wils71}
{Wilson}, R.~E. \& {Devinney}, E.~J. 1971, \apj, 166, 605

\bibitem[{{Yoon} {et~al.}(1994){Yoon}, {Honeycutt}, {Kaitchuck}, \&
  {Schlegel}}]{yoon94}
{Yoon}, T.~S., {Honeycutt}, R.~K., {Kaitchuck}, R.~H., \& {Schlegel}, E.~M.
  1994, \pasp, 106, 239

\bibitem[{{Zejda}(2008)}]{zejd08b}
{Zejda}, M. 2008, Analysis of eclipsing binary light curves: PhD Thesis

\bibitem[{{Zejda} {et~al.}(2008){Zejda}, {Mikul{\'a}{\v s}ek}, \&
  {Wolf}}]{zejd08}
{Zejda}, M., {Mikul{\'a}{\v s}ek}, Z., \& {Wolf}, M. 2008, \aap, 489, 321

\bibitem[{{Zejda} {et~al.}(2006){Zejda}, {Mikul{\'a}{\v s}ek}, {Wolf}, \&
  {Pejcha}}]{zejd06}
{Zejda}, M., {Mikul{\'a}{\v s}ek}, Z., {Wolf}, M., \& {Pejcha}, O. 2006, \apss,
  304, 159

\bibitem[{{Zverko} {et~al.}(2007){Zverko}, {{\v Z}i{\v z}{\v n}ovsk{\'y}},
  {Mikul{\'a}{\v s}ek}, \& {Iliev}}]{zver07}
{Zverko}, J., {{\v Z}i{\v z}{\v n}ovsk{\'y}}, J., {Mikul{\'a}{\v s}ek}, Z., \&
  {Iliev}, I.~K. 2007, Contributions of the Astronomical Observatory Skalnate
  Pleso, 37, 49
\end{thebibliography}
\end{document}